\documentclass[preprint,showpacs,preprintnumbers,amsmath,amssymb]{revtex4}

\usepackage{graphicx}
\usepackage{dcolumn}
\usepackage{bm}
\usepackage{color}
\usepackage{hyperref}
\usepackage{amssymb}
\usepackage{amsmath,empheq}
\usepackage{slashed}

\usepackage{tikz}
\usepackage{bm,bbm}
\usepackage{xcolor}
\usepackage{physics}
\usepackage{caption}
\usepackage{subcaption}

\newcommand{\be}{\begin{eqnarray}}
\newcommand{\ee}{\end{eqnarray}}
\newcommand{\id}{\mathbbm{1}}
\makeatletter
\def\l@subsection#1#2{}
\def\l@subsubsection#1#2{}
\makeatother

\begin{document}

\title{Topological invariant of multilayer Haldane models with irregular stackings}

	\author{Xi Wu}
	\email{wuxi5949@gmail.com, wuxi5949@hnu.edu.cn}
	\affiliation{School of Physics and Electronics, Hunan University, Changsha 410082, China}

\begin{abstract}
We study multilayer Haldane models with irregular type of stacking, considering the nearest interlayer hopping. We prove that the value of the topological invariant is equal to the number of layers times the value of the topological invariant of monolayer Haldane model, regardless of stacking type, and interlayer hoppings do not induce gap closing and phase transitions. 
\end{abstract}

\pacs{}
\maketitle

\tableofcontents

\section{Introduction}
Multilayer graphene have been studied quite a lot in the past two decades and their band structures have been well analyzed with different types of stacking. In nature, stable graphite has Bernal type, which corresponds to ABAB... stacking, rhombohedral type, which correspond to ABCABC...stacking and turbostratic type, which corresponds to irregular stacking that mixes both\cite{Guinea:2006aa,KATSNELSON200720,Min2008,KoshinoMcCann2013}. It is well-known that the type of stacking 
affect the band structure and gives different phenomena in, for instance quantum transport\cite{Min2008,MinMac2008,Bao:2011aa}, optical absorption\cite{Heinz2010,KoshinoMcCann2013}.

On the other hand, Haldane model is constructed also in a hexagonal lattice, the same lattice structure as graphene. Haldane model\cite{Haldane1988} may be the first example of Chern insulator invented, showing the topological structure of electronic Hamiltonian can define new phases of matter. The anomalous Hall conductivity is unchanged by any perturbation if the band gap remains open\cite{Qi2011}. Haldane model thus can be stacked in a similar way to make multilayer Haldane models. Questions naturally arise: Do multilayer Haldane models resemble multilayer graphene in any way? How does the stacking types and the interlayer hopping parameters affect the properties of multilayer Haldane models such as anomalous Hall conductivity? 

For ABC stacking: It was shown analytically\cite{WU2022114863} that, when only the nearest neighbor hopping is taking into account, the Hall conductivity is proportional to the number of layers times the Hall conductivity in monolayer Haldane model and interlayer hopping parameter do not induce band closing. How about other stacking type such as Bernal or even turbostratic?

Moreover, constructing large Chern number device has been an interesting topic in the passed decade. One approach shows that taking into account distant hoppings in a monolayer Haldane model can give rise to larger Chern numbers\cite{Sticlet2013} and this gives beautiful analytical results. Another way is to consider multilayer models but they are usually done in numerical methods\cite{Neupert:2011aa,Trescher:2012aa}. This makes one suspect whether analytical method can be practical in determining the Chern number in multilayer models.

In this paper, generalizing the method in \cite{WU2022114863}, we provide the analytical approach to study the topological number of multilayer Haldane models. Limited to the nearest neighbor interlayer hoppings, we find a one-to-one correspondence between the spectra, eigen-wavefunctions of multilayer graphene and that of multilayer Haldane model. 
Using this correspondence, we prove that the property for ABC stacking is valid for all the stacking types mentioned above and that the values of each interlayer hopping parameters do not matter, either. 

The paper is organized as following: In Sec. (\ref{MugrMuHa}) we review the construction of multilayer graphene and propose multilayer Haldane model in a similar way; in Sec. (\ref{Proof}) we show the correspondence and give the proof; in Sec.(\ref{Eg}) we give examples which violate some of the conditions of the prove, leading to gap closing and thus phase transitions; in Sec. (\ref{Dis}) of discussion, we discuss the gap closing at the limit when interlayer hopping parameters go to infinity, the applications of our result and possible future directions.

\section{Multilayer graphene and Multilayer Haldane model}\label{MugrMuHa}
In Sec.(\ref{Revmugra}), we review the stacking types of multilayer graphene for the benefit of defining our multilayer Haldane model; in Sec.(\ref{MuHad}) we define multilayer Haldane models with irregular stacking and show an important anti-commutation relation that serves for the proof.
\subsection{Review of multilayer graphene}\label{Revmugra}
Here we mainly follow the discussion in \cite{Min2008,MinMac2008}.
Graphene is a two-dimensional set of carbon atoms, arranged into
a honeycomb structure. The structure can be decomposed into
two sublattices, called $\alpha$ and $\beta$.
 Monolayer graphene has linear energy-momentum (dispersion)
relation near the Fermi level 
that the conduction band contacts with the valence band at two Fermi points 
K and K', having effective Hamiltonian
\be
	H_1^G=\left[\begin{array}{cc}
	0 & v(p_1-ip_2) \\
	v(p_1+ip_2) &0 \end{array}\right],
\ee
where $v$ is the Fermi-velocity and $p_1$ $p_2$ are the momenta.

Considering multilayer case, because of the $C_3$ symmetry of
the honeycomb lattice, there are three distinct positions
for the latter layers, 
labelled A-, B-, and C-type.
If one takes into account only the coupling between the nearest layers,
the effective Hamiltonian of multilayer graphene can be written as
\begin{eqnarray}\label{Gn}
	\mathcal{H}^G_{n}=\left[\begin{array}{ccccc}
	H_1^G & \mathfrak{t}_{12}^{T} &  & &
	\\ \mathfrak{t}_{12} & H_1^G & \mathfrak{t}_{23}^{T}  &  &
	\\  &  \mathfrak{t}_{23} & H_1^G & \mathfrak{t}_{34}^{T} &
	\\  &  & &...  &
	\\ &  & & \mathfrak{t}_{n-1,n} &H_1^G \end{array}\right]_{2n\times2n}\,.
\end{eqnarray}
$\mathfrak{t}_{i-1,i}$ is the coupling matrix
between layer $i-1$ and layer $i$, defining the stacking types.
%
%

The stacking
patterns are AA..., ABAB..., and ABCABC... and irregular patterns with mixing between ABAB... and ABCABC....
The simplest case is AA stacking, defined by
\begin{eqnarray}
	\mathfrak{t}_{i-1,i}=t\left[\begin{array}{cc}
	1 & 0 \\
	0 &1 \end{array}\right].
\end{eqnarray}
The $\alpha/\beta$ sublattices from one layer overlaps with the same type of  sublattices of the next layer, making it energetically unstable. The spectrum is 
\be
	E^{\pm}_r=\pm v|\mathbf{p}| +2t\cos(\frac{r\pi}{n+1})\,,r=1,...,n\,.
\ee
The most energetically favorable one is ABAB... stacking. It is defined by
\begin{eqnarray}
	\mathfrak{t}_{2i-1,2i}=t\left[\begin{array}{cc}
	0 & 1 \\
	0 &0 \end{array}\right],
	\quad 
	\mathfrak{t}_{2i,2i+1}=t\left[\begin{array}{cc}
	0 & 0 \\
	1 &0 \end{array}\right],
\end{eqnarray}
and the spectrum is 
\be
	E^{\pm}_r=\pm \sqrt{v^2|\mathbf{p}|^2+t^2\cos^2(\frac{r\pi}{n+1})} +t\cos(\frac{r\pi}{n+1})\,,r=1,...,n\,.
\ee
Another common example is ABC stacking with
\begin{eqnarray}
	\mathfrak{t}_{i-1,i}=t\left[\begin{array}{cc}
	0 & 1 \\
	0 &0 \end{array}\right].
\end{eqnarray}
The exact spectrum is not solved, instead the effective Hamiltonian is given by
\be\label{HABCn}
	H^{eff}_{n}=-\frac{1}{t^{n-1}}
	\left(\begin{array}{cc}
	0 & v^{n}(p_1-ip_2)^{n} 
	\\ v^{n}(p_1+ip_2)^{n} & 0\end{array}\right)\,.
\ee
with dispersion 
\be
	E^{\pm}_n= \pm\frac{v^np^n}{t^{n-1}}\,.
\ee

For an irregular stacking n-layer graphene, we apply the chiral decomposition: (i) Identify the longest ABC stacking chain with its layer, say $J_1$ and partition it out; (ii) repeat step (i) until all the layers are exhausted. Then we have $J_1+J_2+...J_D=n$.
And the effective Hamiltonian of the whole multilayer graphene is
\be
	\mathcal{H}^{eff}_n\approx H_{J_1}\oplus H_{J_2}\oplus...H_{J_{D}}\,, 
\ee
where each $H_{J_i}$ is of form Eq. (\ref{HABCn}).
\subsection{Multilayer Haldane model}\label{MuHad}
In this subsection, we discuss various kinds of multilayer Haldane models corresponding to multilayer graphene models considered in the previous subsection. The monolayer Haldane model has the Hamiltonian
\begin{align}
	H^H_1= h_1 \sigma_1+h_2 \sigma_2+ h_3\sigma_3\,,
\end{align}
here we ignore the term proportional to identity matrix for simplicity, which can be set to zero in the original Haldane model by setting $\cos \phi=0$. 
And the n-layer Haldane model has the Hamiltonian
\be\label{Hn}
	\mathbb{H}^H_{n}=\left[\begin{array}{ccccc}
	H^H_1 & \mathfrak{t}_{12}^{T} &  & &
	\\ \mathfrak{t}_{12} & H^H_1 & \mathfrak{t}_{23}^{T}  &  &
	\\  &  \mathfrak{t}_{23} & H^H_1 & \mathfrak{t}_{34}^{T} &
	\\  &  & &...  &
	\\ &  & & \mathfrak{t}_{n-1,n} &H^H_1\end{array}\right]_{2n\times2n}&=&\sum_ih_i\id_n\otimes\sigma_i+T\nonumber\\
	&=&h_3\id_n\otimes\sigma_3+\mathbb{H}^G_{n}\,,
\ee
where 
\be\label{matht}
	\mathfrak{t}_{i,i+1}=\left[\begin{array}{cc}0 & t_{i,i+1} \\0 & 0\end{array}\right] \text{ or }\left[\begin{array}{cc}0 & 0 \\t_{i,i+1} & 0\end{array}\right]=t_{i,i+1}(\sigma_1\pm i \sigma_2)
\ee
are the inter-layer hopping matrices.  $\mathbb{H}^G_{n}$ have the same matrix structure as, and at K points identical to $\mathcal{H}^G_{n}$. In the following part of this paper, we will refer to $\mathbb{H}^G_{n}$ instead of $\mathcal{H}^G_{n}$ as multilayer graphene.

It is obvious that $h_3\id_n\otimes\sigma_3$ anti-commutes with $h_1\id_n\otimes\sigma_1$ and $h_2\id_n\otimes\sigma_2$. Let's show $\{T,h_3\id_n\otimes\sigma_3\}=0$ in the following: 
\be
	\{\mathfrak{t}_{i,i+1},\sigma_3\}=0 \Leftrightarrow
	\mathfrak{t}_{i,i+1}\sigma_3=-\sigma_3\mathfrak{t}_{i,i+1}
\ee
thus 
\be
	\id_n\otimes \sigma_3 ~T=
-T~\id_n\otimes \sigma_3
\ee
and thus 
\be
	\{T,h_3\id_n\otimes\sigma_3\}=
	0\,.
\ee
As a result, we get
\be\label{h3HGn}
	\{h_3\id_n\otimes\sigma_3, \mathbb{H}^G_{n}\}=0\,,
\ee
and because of this anti-commutation relation, there is a one-to-one correspondence between $\mathbb{H}^G_{n}$ and $\mathbb{H}^H_{n}$.
\section{The one-to-one correspondence and the proof}\label{Proof}
In this section, we prove that for all the stacking types considered in \cite{MinMac2008} except AA stacking(we consider it in Sec. (\ref{AA})), the topological invariant responsible for anomalous Hall conductivity is proportional to the number of layers times that of monolayer Haldane model. The basic logic is the same as in \cite{WU2022114863}, in which we the topological number in the zero interlayer hopping limit and then prove that the gap does not close for any finite values of interlayer hoppings. This section is set into four subsections: in Sec. (\ref{inthop}) We calculate the topological invariant of multilayer Haldane model when the interlaced hoppings are turned off; 
in Sec.(\ref{Lem}) we prove a one-to-one correspondence between multilayer graphene and multilayer Haldane model, which shows that the gap closing of multilayer Haldane model cannot happen in Brillouin zone except at multilayer graphene gap closing point; 
 in Sec.(\ref{Gapmugr}) we study the gap closing condition of multilayer graphene and find that interlayer hoppings do not change it; in Sec.(\ref{Fproof}) we summarize the ingredients in previous subsections and explains that 
the gap of multilayer Haldane model is protected by monolayer Haldane model gap and together with the result in Section. (\ref{inthop}) we complete the proof.
\subsection{Interlayer hopping zero limit}\label{inthop}
\label{topinv}
The result in this subsection is not new, but for the completeness of the proof we put it here. The topological invariant \cite{ISHIKAWA1987523,Golterman1992ub,Volovik_2009,ZUBKOV2016298} responsible for the conductivity of anomalous quantum Hall effect, is defined as follows
\be\label{TIN}
	\mathcal{N}[\mathbb{G}]=\frac{1}{3!}\int \frac{d^3 p}{(2\pi)^2} \,\epsilon^{ijk}\Tr(\mathbb{G}\partial_i \mathbb{G}^{-1}\mathbb{G}\partial_j \mathbb{G}^{-1}\mathbb{G}\partial_k \mathbb{G}^{-1})\,.
\ee
In these expressions $\mathbb{G}$ is the two-point Green function of electrons. In this section, we consider the limit that the interlayer hopping is zero, namely $t_{\bot}=0$. $\mathbb{G}$ is expressed as
\be
	\mathbb{G}_{}^{-1}=i\omega-\mathbb{H}^H_{n}|_{t_\bot=0}=\left[\begin{array}{cccc}
	Q_1 & &  & 
	\\  & Q_2 &   &  
	\\  &   &...  &
	\\ &  &  &Q_n\end{array}\right]_{2n\times2n}\,. 	
\ee
where $Q_i=i\omega-H_1^i$ and $i$ is the label of the layer.
The matrix $\mathbb{G}^{-1}$ becomes a direct sum of  $Q_i$ . If a matrix $\mathbb{G}$ is a direct sum of the two other matrices $G_1$ and $G_2$, then the topological invariant, or the winding number, of $\mathbb{G}$ will be the sum of the topological invariant of $G_1$ and that of $G_2$.
 Namely, if $\mathbb{G}=\left[\begin{array}{cc} G_1 & 0 \\0 & G_2\end{array}\right]$,  then $N[\mathbb{G}]=N[G_1]+N[G_2]$  \cite{nakahara2003geometry}.Because: 
\be
	\mathcal{N}[\mathbb{G}]&=&\frac{1}{3!}\int \frac{d^3 p}{(2\pi)^2}\,\epsilon^{ijk} \Tr(\mathbb{G}\partial_i \mathbb{G}^{-1}\mathbb{G}\partial_j \mathbb{G}^{-1}\mathbb{G}\partial_k \mathbb{G}^{-1})\nonumber \\
	&=&\frac{1}{3!}\int \frac{d^3 p}{(2\pi)^2}\, \epsilon^{ijk}\Tr(G_1\partial_i G^{-1}_1G_1\partial_j G^{-1}_1G_1\partial_k G^{-1}_1+G_2\partial_i G^{-1}_2G_2\partial_j G^{-1}_2G_2\partial_k G^{-1}_2)\nonumber \\
	&=&\mathcal{N}[G_1]+\mathcal{N}[G_2]\,.
\ee
Therefore,
\be\label{NGn}
	\mathcal{N}[\mathbb{G}_{n}]=\sum_i\mathcal{N}[G_i]\,.
\ee
One can see that without the inter-layer hoppings, the n-layer Haldane model would have the topological invariant with the value equal to the sum of the topological invariants of each layer. This would be generalized if we turn on the inter-layer hopping and the energy gap does not close, which will be shown in the next subsection.

\subsection{One-to-one correspondence}\label{Lem}
In this subsection we prove a lemma that tells there is a one-to-one correspondence of the energy spectrum and eigenfunction between multilayer graphene and multilayer Haldane model. 

 We start from considering matrices
\be\label{HHGH3}
	H_H=H_G+H_3\,,
\ee
satisfying
\be\label{HGH3}
	\{H_G,H_3\}=0,
\ee
and 
\be\label{H3h3}
	H_3^2=h_3^2~ \id\,,
\ee
where $h_3$ is some number. These conditions are the ones satisfied by the Hamiltonians of multilayer graphene and multilayer Haldane model, shown in Sec.(\ref{MuHad}). 

The lemma is as following:
For each eigenequation
\be\label{HGE}
	(H_G-E_i^G)\psi_i^G=0\,,
\ee
there exists an eigen-equation
\be\label{HHE}
	(H_H-E_i^H)\psi_i^H=0\,,
\ee
satisfying
\be\label{EH2EG2}
	(E_i^H)^2=h_3^2+(E_i^G)^2\,,
\ee
and vice versa.

Proof: From Eq. (\ref{HGE}) we get
\be\label{HG2EG2}
	\Big(H_G^2-(E_i^G)^2\Big)\psi_i^G=0\,.
\ee
Eq. (\ref{HHGH3}) (\ref{HGH3})and (\ref{H3h3}) gives
\be\label{HH2HG2}
	H_H^2=H_G^2+h_3^2~\id \,.
\ee
Substituting Eq. (\ref{HH2HG2}) into Eq. (\ref{HG2EG2}), we have
\be\label{HH2hE}
	\Big(H_H^2-h_3^2-(E_i^G)^2\Big)\psi_i^G=0\,.
\ee
There are two solutions for Eq. (\ref{HH2hE})
\be
	\Big(H_H\mp\sqrt{h_3^2+(E_i^G)^2}\Big)\psi_i^{H\pm}=0\,.,
\ee
with 
\be
	\psi_i^{H\pm}=\Big(H_H\pm\sqrt{h_3^2+(E_i^G)^2}\Big)\psi_i^G\,.
\ee
Define
\be
\label{EHh3EG}
	E_i^H&:=&\text{sgn}(E_i^G) \sqrt{h_3^2+(E_i^G)^2}\\
\label{pHh3pG}	\psi_i^{H}&:=&\Big(H_H+\text{sgn}(E_i^G)\sqrt{h_3^2+(E_i^G)^2}\Big)\psi_i^G\,,
\ee
we arrive at Eq. (\ref{HHE}) and (\ref{EH2EG2}). The sign of $E_i^H$ is chosen such that at the limit
$h_3\to0$
Eq. (\ref{HGE}) and (\ref{HHE})become identical, which means
\be
	E_i^H(h_3\to 0)\to E_i^G\,,
\ee
and also $\psi_i^{H}\ne0$.
The above procedure can be reversed: starting from Eq. (\ref{HHE}) we can derive Eq. (\ref{HGE}) and (\ref{EH2EG2}) so the lemma is proven. 

Remark: the condition Eq. (\ref{H3h3}) makes sure that the correspondence is one-to-one, without it, there is still a correspondence similar as Eq. (\ref{EHh3EG}) and (\ref{pHh3pG})but not one-to-one, between the two spectra, with $h_3^2$ replaced by eigenvalues of $H_3^2$.
\subsection{Gap closing condition of multilayer graphene}\label{Gapmugr}
Eq. (\ref{EHh3EG}) tells us that the gap closing condition of multilayer Haldane model, namely $E^H_i=0$ for some $i$  is that $h_3=E^G_i=0$. In this subsection, we look for gap closing condition of multilayer graphene $E^G_i=0$. The secular determinant of multilayer graphene is
\be\label{detHnG}
	&&\det(\mathbb{H}^G_{n}-E^G \id_n)=\det(h_1\id_n\otimes \sigma_1+h_2\id_n\otimes \sigma_2+T-E^G\id)
\ee
and the gap closes at half-filling when 
\be\label{detHnG0}
	\det(\mathbb{H}^G_{n}-E^G\id_n)\Big|_{E^G=0}=\det(\mathbb{H}^G_{n})=0\,.
\ee
The choices of $h_1$ and $h_2$ that satisfy Eq. (\ref{detHnG0}) determine the gap closing points. Though stacking type and values of $t_{i.i+1}$ in general determine the energy eigenvalues of multilayer graphene and it is very hard or  maybe impossible to find the energy spectrum, Eq. (\ref{detHnG0}) is doable. 

Next we show that 
\be
	\det(\mathbb{H}^G_{n})=(-h_1^2-h_2^2)^n
\ee
for $\mathfrak{t}_{i,i+1}$ in Eq. (\ref{matht}) and any value of $t_{i.i+1}\,,i=1,...,n-1$. When $t_{i.i+1}=0$ for all $i=1,...,n-1$ it is easy to see that
\be
	\det(\mathbb{H}^G_{n})=(-h_1^2-h_2^2)^n\,,
\ee
and we just need to prove terms involving $t_{i.i+1}$ do not contribute to the determinant.  We use mathematical induction. From here we change the notation of $\mathfrak{t}_{i,i+1}$ from $t_{i,i+1}$ into $\left[\begin{array}{cc}0 & u_i \\v_i & 0\end{array}\right]$ for convenience. For $n=2$ we have 
\be
	\det(\mathbb{H}^G_{2})
	&=&
	\left|\begin{array}{cccc}
	0 & h_1-ih_2 & &v  \\h_1+ih_2  & 0 & u &  \\ & u & 0 & h_1-ih_2 \\ v&  & h_1+ih_2 & 0
	\end{array}\right|=(-h_1^2-h_2^2)^2\,,
\ee
for either $u=0$ or $v=0$. Suppose for $n=k$
\be
	\det(\mathbb{H}^G_{k})=(-h_1^2-h_2^2)^k\,,
\ee
we calculate $\det(\mathbb{H}^G_{k+1})$. It is expressed as following:
\be
	&&\det(\mathbb{H}^G_{k+1})\nonumber\\
	&=&\left|\begin{array}{ccccccc}
	0 & h_1-ih_2 & 0 & v_1 & 0 & 0 & \mathbf{0} 
	\\h_1+ih_2 & 0 & u_1 & 0 & 0 & 0 & \mathbf{0} 
	\\0 & u_1 & 0 & h_1-ih_2 & 0 & v_2 & \mathbf{0} 
	\\v_1 & 0 & h_1+ih_2& 0 & u_2 & 0 & \mathbf{0} 
	\\0 & 0 & 0 & u_2 & 0 & h_1-ih_2 & ... 
	\\0 & 0 & v_2 & 0 & h_1+ih_2 & 0 & ... 
	\\\mathbf{0} & \mathbf{0} & \mathbf{0} & \mathbf{0} & ... & ... & ...\end{array}\right|\,.
\ee
Let us consider $u_1=0$ and $v_1=0$ separately. $u_1=0$ or $v_1=0$ can contribute to the determinant only if their cofactors are nonzero. If $u_1=0$, we can show that the cofactors of the two $v_1$s equal to zero:
\be
	A_{4,1}=-\left|\begin{array}{cccccc}
	h_1+ih_2 & 0 & 0 &  0 & 0 & \mathbf{0} 
	\\0 & 0 & 0 &  0 & v_2 & \mathbf{0} 
	\\v_1 & 0 & h_1+ih_2&  u_2 & 0 & \mathbf{0} 
	\\0 & 0 & 0 &  0 & h_1-ih_2 & ... 
	\\0 & 0 & v_2 &  h_1+ih_2 & 0 & ... 
	\\\mathbf{0} & \mathbf{0} & \mathbf{0} &  ... & ... & ...\end{array}\right|=0
\ee
because the second column is zero and similarly $A_{1,4}$ is zero because its second row is zero.
If $v_1=0$, the cofactors of the two  $u_1$s are zero:
\be
	A_{2,3}=-\left|\begin{array}{cccccc}
	0 & h_1-ih_2 & 0  & 0 & 0 & \mathbf{0} 
	\\0 & u_1 & h_1-ih_2 & 0 & v_2 & \mathbf{0} 
	\\0 & 0 &  0 & u_2 & 0 & \mathbf{0} 
	\\0 & 0 &  u_2 & 0 & h_1-ih_2 & ... 
	\\0 & 0 &  0 & h_1+ih_2 & 0 & ... 
	\\\mathbf{0} & \mathbf{0} & \mathbf{0} &  ... & ... & ...\end{array}\right|\,.
\ee	
because the first column is zero and similarly $A_{3,2}$ is zero because its first row is zero. Therefore there is no contribution from $u_1$ or $v_1$. As a result,
\be
	\det(\mathbb{H}^G_{k+1})=(-h_1^2-h_2^2)\det(\mathbb{H}^G_{k})=(-h_1^2-h_2^2)^{k+1}\,.
\ee
Therefore, the gap closing condition is still $h_1=h_2=0$ unchanged by interlayer hoppings. Remark: this result can be generalized, namely the $h_1$ and $h_2$ in each layer can be different $h^i_1$ and $h^i_2$, then we will have  $\det(\mathbb{H}^G_{n})=\prod_{i=1}^{n} (-(h_1^i)^2-(h_2^i)^2)$ and the gap closing condition is $h_1^i=h_2^i=0$ for one of the $i$'s.
\subsection{The final proof}\label{Fproof}
Let us summarize the conclusions in precious subsections: from Sec.(\ref{MuHad}) the Hamiltonian of multilayer Haldane model composes of $h_3\id_n\otimes\sigma_3$ and the Hamiltonian of multilayer graphene, which anti-commute with each other, and from Sec. (\ref{Lem}) this leads to a one-to-one correspondence between the spectra of them Eq. (\ref{EHh3EG}). As far as $AA$ stacking is excluded,  we showed in Sec (\ref{Gapmugr}) multilayer graphene band gap closes, namely $E_i^G=0$, only when 
$h_1=h_2=0$, but then $h_3\ne0$ as is required by the gap of monolayer Haldane model:
\be
	h_1^2+h_2^2+h_3^2>0\,.
\ee
Therefore  $h_3=0$ and $E_i^G=0$  can never be satisfied simultaneously except the possibility at the limit $t_{i,i+1} \to \infty$. So the topological invariant of multilayer Haldane model is not modified by the introduction of the nearest interlayer hoppings, even for the irregular stacking types. Sec. (\ref{inthop}) shows that at the limit interlayer hoppings are zero, the topological invariant of multilayer Haldane model is the sum of all topological invariant of each monolayer Haldane models.
This completes the proof. In a sense, our proof is not even limited to Haldane model, because we do not need the detailed information of $h_1$ $h_2$ and $h_3$. 

Remark: This conclusion can be generalized into the situation that for each layer with label $i$, $h_1^{i}$ and $h_2^i$ do not have to be the same with other layers. The bottom line is that $h_3^i$ has to be the same in order to anti-commute with $T$ matrix. Since each layer is an insulator, $(h_1^i)^2+(h_2^i)^2+h_3^2>0$. Therefore there is no phase transition when turning on any of the interlayer hopping. Then the topological invariant of the whole layered system is  the sum of the topological invariants of each layer.  

\section{Examples of models that do have phase transitions}\label{Eg}
To show the non-triviality of the result in the previous section, we consider two examples that do have phase transitions, violating one of the requirements in previous sections. For simplicity we only consider bilayer models. The models are: AA stacking, in which Eq. (\ref{h3HGn}) is not obeyed, and $\alpha\beta\&\beta\alpha$ stacking, in which $h_3=0$ and $E_i^G=0$ can be simultaneously satisfied. 

\subsection{AA stacking}\label{AA}
The AA stacking model is defined as following:
\be\label{H2AA}
	H^2_{AA}&=&
	\left[\begin{array}{cccc}
	h_3 & h_1-ih_2 &t &  
	\\h_1+ih_2  & -h_3 &  &  t
	\\ t&  & h_3 & h_1-ih_2 
	\\ &t  & h_1+ih_2 & -h_3
	\end{array}\right]\nonumber
	\\
	&=:&\sum_{i=1}^3 h_i \id \otimes \sigma_i+t\sigma_1\otimes\id\,.
\ee
$H^2_{AA}$ is block-diagonalized by a transformation $S=\frac{1}{\sqrt2}(\id-i\sigma_2)\otimes\id$:
\be
	S H^2_{AA} S^{-1}=\left[\begin{array}{cccc}
	h_3+t & h_1-ih_2 & &  
	\\h_1+ih_2  & -h_3+t &  &  
	\\ &  & h_3-t & h_1-ih_2 
	\\ &  & h_1+ih_2 & -h_3-t
	\end{array}\right]
\ee
therefore the spectrum is 
\be
	E_{AA}=\pm t\pm\sqrt{\sum_ih_i^2}=:\pm t\pm h\,.
\ee
The band-closing points are 
\be
	|t|=h_{min},h_{max}\,
\ee
and if we tune the value of $t$ from zero to infinity the bilayer model will transit from topological/normal(depending on the monolayer model) insulator to metal and then to normal insulator.

 \subsection{$\alpha\beta\&\beta\alpha$ stacking}

We define $\alpha\beta\&\beta\alpha$ stacking as following: there are two interlayer hoppoings: both from the $\alpha$ sub degrees of freedom to the $\beta$ sub degrees of freedom and from the $\beta$ sub degrees of freedom to the $\alpha$ sub degrees of freedom. The continuum models were introduced in \cite{ZhangMacDonald2013,Hashimoto2016}. The Hamiltonian is as following
\be\label{H2abba}
	H^2_{\alpha\beta\&\beta\alpha}&=&
	\left[\begin{array}{cccc}
	h_3 & h_1-ih_2 & &v  \\h_1+ih_2  & -h_3 & u &  \\ & u & h_3 & h_1-ih_2 \\ v&  & h_1+ih_2 & -h_3
	\end{array}\right]\nonumber
	\\
	&=&\sum_{i=1}^3 h_i \id \otimes \sigma_i+\frac{u+v}{2}\sigma_1\otimes\sigma_1+\frac{u-v}{2}\sigma_2\otimes\sigma_2
	\nonumber
	\\
	&=:&\sum_{i=1}^3 h_i \id \otimes \sigma_i+t_1\sigma_1\otimes\sigma_1+t_2\sigma_2\otimes\sigma_2\,.
\ee
When $u=0$ or $v=0$, this Hamiltonian reduces to bilayer case of the model in Eq. (\ref{Hn}). Although the gap does not close for the model in Eq. (\ref{Hn}), for general $u$ and $v$, as we will see, the gap does close. 
\begin{figure}[th]
 \begin{center}
    \begin{subfigure}[b]{0.3\textwidth}
         \includegraphics[width=\textwidth]{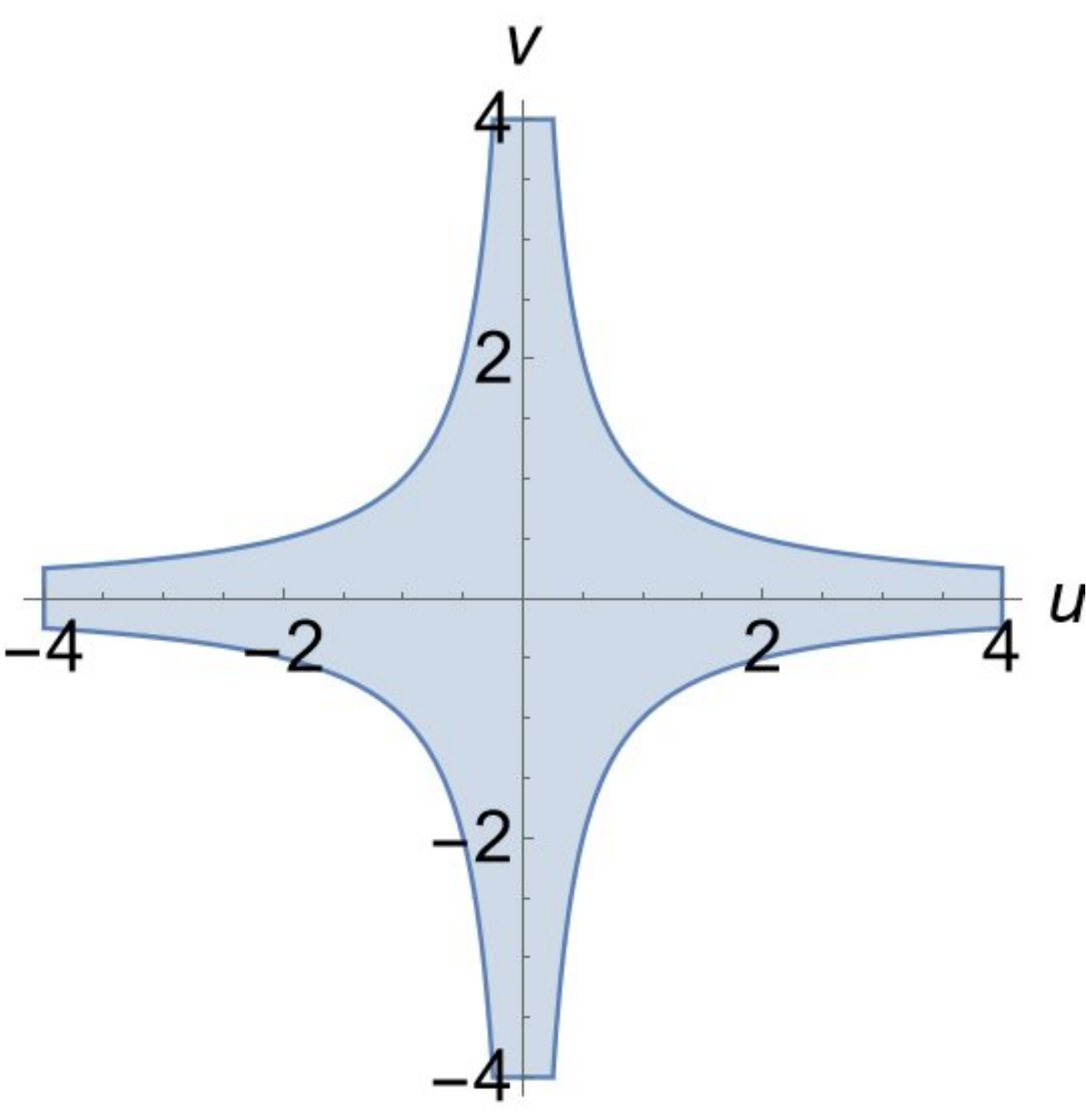}
         \caption{$m = 1$, and the $x$,$y$ axises are respectively $u$ and $v$}
         \label{fig:d1}
     \end{subfigure}
    \hfill
    \begin{subfigure}[b]{0.3\textwidth}
         \includegraphics[width=\textwidth]{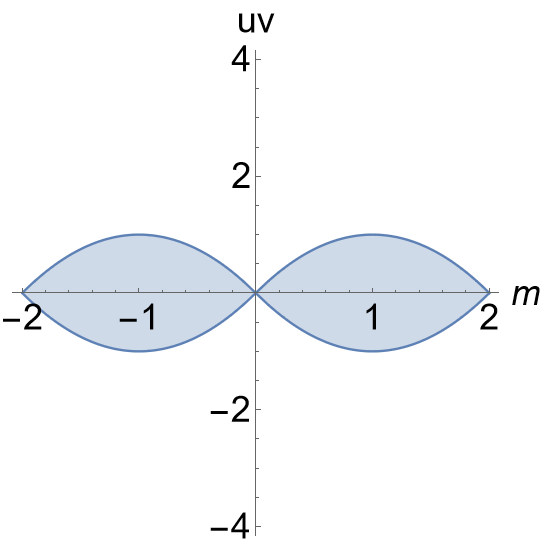}
         \caption{The $x$,$y$ axises are respectively $uv$ and $m$}
         \label{fig:d2}
     \end{subfigure}
    \hfill    
    \begin{subfigure}[b]{0.3\textwidth}
         \includegraphics[width=\textwidth]{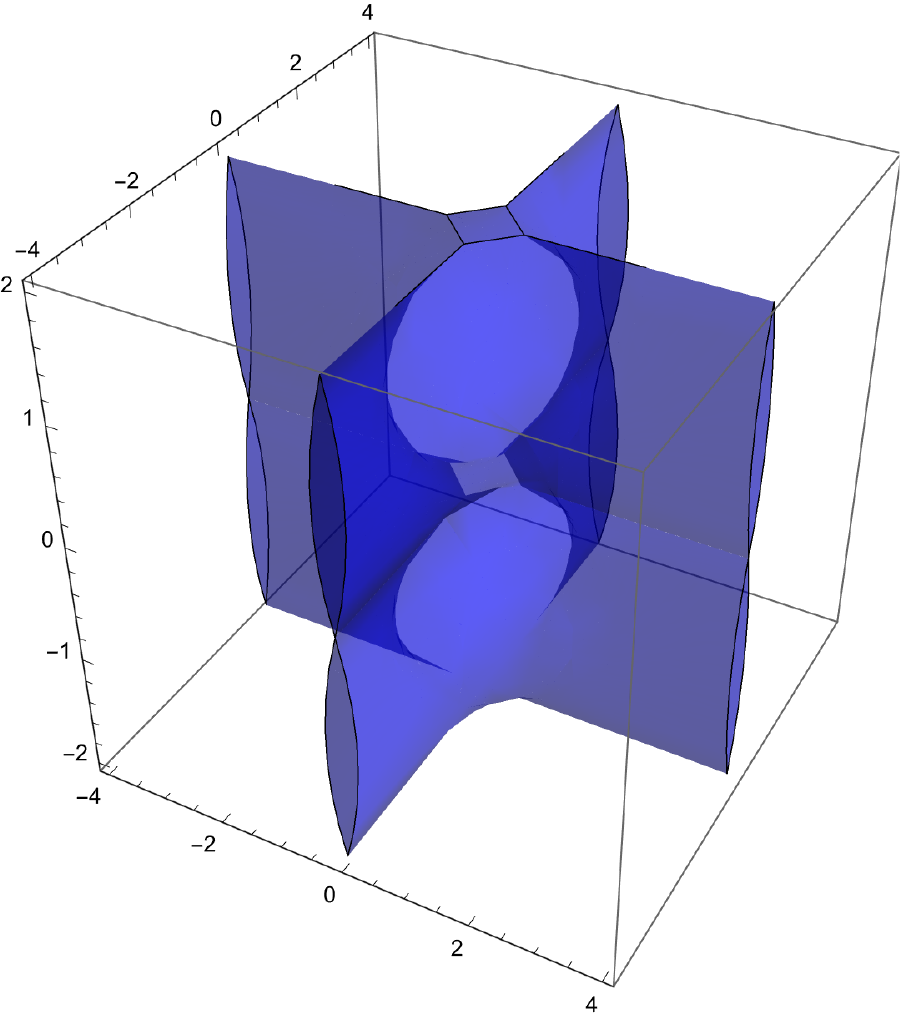}
         \caption{Choosing the $x$,$y$, $z$ axis to be  $u$, $v$ and $m$}
         \label{fig:d3}
     \end{subfigure}
  \end{center}
 \caption{\raggedright The phase diagram of  $H^2_{\alpha\beta\&\beta\alpha}$. The colored regions are topological nontrivial if monolayer Hamiltonian is so.}
 \label{fig:quenchABBAH}
 \end{figure}
The spectrum of this Hamiltonian satisfies
\be
	E_{\alpha\beta\&\beta\alpha}^2=h_3^2+h_1^2+h_2^2+t_1^2+t_2^2\pm2\sqrt{h_1^2t_1^2+h_2^2t_2^2+t_1^2t_2^2}\,,.
\ee
Now we define the following function
\be
	f&:=&(h_1^2+h_2^2+t_1^2+t_2^2)^2-4(h_1^2t_1^2+h_2^2t_2^2+t_1^2t_2^2)\,,
\ee
and then gap closing condition can be reinterpreted as 
\be
	 \left\{ \begin{array}{rcl}
       &f=0
         \\
        &h_3=0\,.
                \end{array}\right.
\ee
$f$ is simplified as
\be
	f=(h_1^2-h_2^2-t_1^2+t_2^2)^2+4h_1^2h_2^2\,.
\ee
So the gap closing condition becomes
\be \label{GCCHBA}
	 \left\{ \begin{array}{rcl}
       &&t_1^2-t_2^2=h_1^2-h_2^2 \text{ or } uv=h_1^2-h_2^2 
         \\  &&h_1h_2=0
         \\
        &&h_3=0\,.
                \end{array}\right.
\ee
In the case of bilayer model of Eq. (\ref{Hn}), all three equations cannot be simultaneously satisfied because $uv=0$. So $uv=0$ case are topologically nontrivial, if the monolayer model is topological. There is a phase transition from topologically nontrivial phase into topologically trivial phase when turning $|uv|$ from zero to infinity: From Eq. (\ref{GCCHBA})we can see when increasing $|uv|$  there are always some values of $|uv|$, say $|uv|_0 $ to close the gap, and after all the gap-closing values, the phase becomes trivial because it is smoothly connected to the case that $u\to \pm\infty$ and $v\to \pm\infty$.

Next let's consider Wilson fermion\cite{Wilson_1977} for each layer of Eq. (\ref{H2abba}) instead of Haldane model for simplicity and study the phase transition. It was also proposed Qi, Wu and Zhang to be realized as spin Hall effect in two-dimensional paramagnetic semiconductors\cite{QWZ2006}. 
The Hamiltonian for Wilson fermion is 
\be
	H_W= (m+\cos p_1+\cos p_2)\sigma_3+\sin p_1 \sigma_1+\sin p_2\sigma_2\,,
\ee
from Eq. (\ref{GCCHBA}) the gap closing condition is
\be
	&&\text{If } uv>0  \left\{ \begin{array}{rcl}
       &&uv=\sin^2 p_1
         \\  &&\sin p_2=0
         \\ &&m\pm1+\cos p_1=0\,.
                \end{array}\right.
         \Rightarrow uv=1-(1\pm m)^2\,, \label{uv>0}
         \\
       &&  \text{If } uv<0  \left\{ \begin{array}{rcl}
       &&uv=-\sin^2 p_2
         \\  &&\sin p_1=0
         \\ &&m\pm1+\cos p_2=0\,.
                \end{array}\right.
         \Rightarrow -uv=1-(1\pm m)^2\,, \label{uv<0}
\ee
in which + or- is taken such that $(1\pm m)^2\le 1$. Eq. (\ref{uv>0}) and  (\ref{uv>0}) can be combined into one condition
\be
	(uv)^2=(1-(1\pm m)^2)^2\,.
\ee

\section{Discussion}\label{Dis}
Though we proved that the gap does not closed for any finite value of interlayer hopping $t_{i,i+1}$, the gap does get smaller and smaller when we increase the value of $t_{i,i+1}$, and closes at $t_{i,i+1}=\infty$. The reason is that at $t_{i,i+1}=\infty$, there is one $E^G$ goes to the zero limit  for all values of $h_1$ and $h_2$ so the gap closing point of multilayer Haldane model is determined solely by $h_3=0$. In \cite{WU2022114863} we showed this for bilayer model. Now we show it is true in n-layer model also, at least  when all $t_{i,i+1}$ have the same value $t_{bot}$. When we increase the value of $t_{bot}$ such that $t_{bot}>>\sqrt{h_1^2+h_2^2}$, then the low energy effective theory nearby $K$ points in \cite{Min2008} becomes a good approximation throughout the Brillouin zone. And we have
\be
	\mathcal{H}^G_n\approx H_{J_1}\oplus H_{J_2}\oplus...H_{J_{D}}\,, 
\ee
where
\be
	H_{J_i}=-\frac{1}{t^{J_i-1}_{\bot}}
	\left(\begin{array}{cc}
	0 & (h_1-ih_2)^{J_i} 
	\\ (h_1+ih_2)^{J_i} & 0\end{array}\right)
\ee
where $J_i$ are integer numbers that corresponding to the chiral decomposition and $J_1+J_2+...J_D=n$. And the energy eigenvalues
\be
	E_{J_i}=\pm\frac{\sqrt{h_1^2+h_2^2}^{J_i}}{t_{\bot}^{J_i-1}} \to 0 \text{ as } t_{\bot} \to 0
\ee 
for $J_i\geq2$. But it is impossible that all $J_i=1$ because of the mechanism of chiral decomposition. Therefore if $t_{bot}$ becomes too big, though it is impossible to transit into another topological/trivial insulator phase, the material becomes a semiconductor practically.

From the proof we can see that multilayer Haldane models even with irregular stacking can be used as a method to build large Chern number device since the phase is understood. Moreover, from Sec. (\ref{Fproof}) ,as the behavior is predictable when changing $h_1^i$ and $h_2^i$ in each layer, we can change the Chern number not only by adding or removing layers but also by modifying each layer. And one way to do this is to consider distant neighbor intra-layer hoppings: as shown in\cite{Sticlet2013}, distant neighbor hoppings can also create large Chern number so the combination of both methods can give even more structures. 

In his paper, we have found the one-to-one correspondence between multilayer graphene and multilayer Haldane models, there may be more interesting phenomena of multilayer Haldane models to be discovered. We only consider nearest neighbor interlayer hoppings. As shown in multilayer graphene\cite{KoshinoMacCann2009}, distant interlayer hoppings can induce trigonal warping in the band structure. It is interesting to understand what will happen to multilayer Haldane model if we allow distant interlayer hoppings. Maybe the one-to-one correspondence will be modified or even ruined. We leave this as a future direction.

 \begin{acknowledgments} X. Wu is grateful for valuable discussions with C.X.Zhang.
\end{acknowledgments}


\bibliography{MultilayerHaldanemodelgeneralizedNotes}

\begin{thebibliography}{23}
\expandafter\ifx\csname natexlab\endcsname\relax\def\natexlab#1{#1}\fi
\expandafter\ifx\csname bibnamefont\endcsname\relax
  \def\bibnamefont#1{#1}\fi
\expandafter\ifx\csname bibfnamefont\endcsname\relax
  \def\bibfnamefont#1{#1}\fi
\expandafter\ifx\csname citenamefont\endcsname\relax
  \def\citenamefont#1{#1}\fi
\expandafter\ifx\csname url\endcsname\relax
  \def\url#1{\texttt{#1}}\fi
\expandafter\ifx\csname urlprefix\endcsname\relax\def\urlprefix{URL }\fi
\providecommand{\bibinfo}[2]{#2}
\providecommand{\eprint}[2][]{\url{#2}}

\bibitem[{\citenamefont{Guinea et~al.}(2006)\citenamefont{Guinea, Castro~Neto,
  and Peres}}]{Guinea:2006aa}
\bibinfo{author}{\bibfnamefont{F.}~\bibnamefont{Guinea}},
  \bibinfo{author}{\bibfnamefont{A.~H.} \bibnamefont{Castro~Neto}},
  \bibnamefont{and} \bibinfo{author}{\bibfnamefont{N.~M.~R.}
  \bibnamefont{Peres}}, \bibinfo{journal}{Phys. Rev. B}
  \textbf{\bibinfo{volume}{73}}, \bibinfo{pages}{245426}
  (\bibinfo{year}{2006}),
  \urlprefix\url{https://link.aps.org/doi/10.1103/PhysRevB.73.245426}.

\bibitem[{\citenamefont{Katsnelson}(2007)}]{KATSNELSON200720}
\bibinfo{author}{\bibfnamefont{M.~I.} \bibnamefont{Katsnelson}},
  \bibinfo{journal}{Materials Today} \textbf{\bibinfo{volume}{10}},
  \bibinfo{pages}{20 } (\bibinfo{year}{2007}), ISSN \bibinfo{issn}{1369-7021},
  \urlprefix\url{http://www.sciencedirect.com/science/article/pii/S1369702106717886}.

\bibitem[{\citenamefont{Min and MacDonald}(2008{\natexlab{a}})}]{Min2008}
\bibinfo{author}{\bibfnamefont{H.}~\bibnamefont{Min}} \bibnamefont{and}
  \bibinfo{author}{\bibfnamefont{A.~H.} \bibnamefont{MacDonald}},
  \bibinfo{journal}{Phys. Rev. B} \textbf{\bibinfo{volume}{77}},
  \bibinfo{pages}{155416} (\bibinfo{year}{2008}{\natexlab{a}}),
  \urlprefix\url{https://link.aps.org/doi/10.1103/PhysRevB.77.155416}.

\bibitem[{\citenamefont{Koshino and McCann}(2013)}]{KoshinoMcCann2013}
\bibinfo{author}{\bibfnamefont{M.}~\bibnamefont{Koshino}} \bibnamefont{and}
  \bibinfo{author}{\bibfnamefont{E.}~\bibnamefont{McCann}},
  \bibinfo{journal}{Phys. Rev. B} \textbf{\bibinfo{volume}{87}},
  \bibinfo{pages}{045420} (\bibinfo{year}{2013}),
  \urlprefix\url{https://link.aps.org/doi/10.1103/PhysRevB.87.045420}.

\bibitem[{\citenamefont{Min and MacDonald}(2008{\natexlab{b}})}]{MinMac2008}
\bibinfo{author}{\bibfnamefont{H.}~\bibnamefont{Min}} \bibnamefont{and}
  \bibinfo{author}{\bibfnamefont{A.~H.} \bibnamefont{MacDonald}},
  \bibinfo{journal}{Progress of Theoretical Physics Supplement}
  \textbf{\bibinfo{volume}{176}}, \bibinfo{pages}{227}
  (\bibinfo{year}{2008}{\natexlab{b}}), ISSN \bibinfo{issn}{0375-9687},
  \eprint{https://academic.oup.com/ptps/article-pdf/doi/10.1143/PTPS.176.227/5322668/176-227.pdf},
  \urlprefix\url{https://doi.org/10.1143/PTPS.176.227}.

\bibitem[{\citenamefont{Bao et~al.}(2011)\citenamefont{Bao, Jing, Velasco, Lee,
  Liu, Tran, Standley, Aykol, Cronin, Smirnov et~al.}}]{Bao:2011aa}
\bibinfo{author}{\bibfnamefont{W.}~\bibnamefont{Bao}},
  \bibinfo{author}{\bibfnamefont{L.}~\bibnamefont{Jing}},
  \bibinfo{author}{\bibfnamefont{J.}~\bibnamefont{Velasco}},
  \bibinfo{author}{\bibfnamefont{Y.}~\bibnamefont{Lee}},
  \bibinfo{author}{\bibfnamefont{G.}~\bibnamefont{Liu}},
  \bibinfo{author}{\bibfnamefont{D.}~\bibnamefont{Tran}},
  \bibinfo{author}{\bibfnamefont{B.}~\bibnamefont{Standley}},
  \bibinfo{author}{\bibfnamefont{M.}~\bibnamefont{Aykol}},
  \bibinfo{author}{\bibfnamefont{S.~B.} \bibnamefont{Cronin}},
  \bibinfo{author}{\bibfnamefont{D.}~\bibnamefont{Smirnov}},
  \bibnamefont{et~al.}, \bibinfo{journal}{Nature Physics}
  \textbf{\bibinfo{volume}{7}}, \bibinfo{pages}{948} (\bibinfo{year}{2011}),
  \urlprefix\url{https://doi.org/10.1038/nphys2103}.

\bibitem[{\citenamefont{Mak et~al.}(2010)\citenamefont{Mak, Shan, and
  Heinz}}]{Heinz2010}
\bibinfo{author}{\bibfnamefont{K.~F.} \bibnamefont{Mak}},
  \bibinfo{author}{\bibfnamefont{J.}~\bibnamefont{Shan}}, \bibnamefont{and}
  \bibinfo{author}{\bibfnamefont{T.~F.} \bibnamefont{Heinz}},
  \bibinfo{journal}{Phys. Rev. Lett.} \textbf{\bibinfo{volume}{104}},
  \bibinfo{pages}{176404} (\bibinfo{year}{2010}),
  \urlprefix\url{https://link.aps.org/doi/10.1103/PhysRevLett.104.176404}.

\bibitem[{\citenamefont{Haldane}(1988)}]{Haldane1988}
\bibinfo{author}{\bibfnamefont{F.~D.~M.} \bibnamefont{Haldane}},
  \bibinfo{journal}{Phys. Rev. Lett.} \textbf{\bibinfo{volume}{61}},
  \bibinfo{pages}{2015} (\bibinfo{year}{1988}),
  \urlprefix\url{https://link.aps.org/doi/10.1103/PhysRevLett.61.2015}.

\bibitem[{\citenamefont{Qi and Zhang}(2011)}]{Qi2011}
\bibinfo{author}{\bibfnamefont{X.-L.} \bibnamefont{Qi}} \bibnamefont{and}
  \bibinfo{author}{\bibfnamefont{S.-C.} \bibnamefont{Zhang}},
  \bibinfo{journal}{Rev. Mod. Phys.} \textbf{\bibinfo{volume}{83}},
  \bibinfo{pages}{1057} (\bibinfo{year}{2011}),
  \urlprefix\url{https://link.aps.org/doi/10.1103/RevModPhys.83.1057}.

\bibitem[{\citenamefont{Wu et~al.}(2022)\citenamefont{Wu, Zhang, and
  Zubkov}}]{WU2022114863}
\bibinfo{author}{\bibfnamefont{X.}~\bibnamefont{Wu}},
  \bibinfo{author}{\bibfnamefont{C.}~\bibnamefont{Zhang}}, \bibnamefont{and}
  \bibinfo{author}{\bibfnamefont{M.}~\bibnamefont{Zubkov}},
  \bibinfo{journal}{Solid State Communications} \textbf{\bibinfo{volume}{353}},
  \bibinfo{pages}{114863} (\bibinfo{year}{2022}), ISSN
  \bibinfo{issn}{0038-1098},
  \urlprefix\url{https://www.sciencedirect.com/science/article/pii/S0038109822001922}.

\bibitem[{\citenamefont{Sticlet and Pi\'echon}(2013)}]{Sticlet2013}
\bibinfo{author}{\bibfnamefont{D.}~\bibnamefont{Sticlet}} \bibnamefont{and}
  \bibinfo{author}{\bibfnamefont{F.}~\bibnamefont{Pi\'echon}},
  \bibinfo{journal}{Phys. Rev. B} \textbf{\bibinfo{volume}{87}},
  \bibinfo{pages}{115402} (\bibinfo{year}{2013}),
  \urlprefix\url{https://link.aps.org/doi/10.1103/PhysRevB.87.115402}.

\bibitem[{\citenamefont{Titus~Neupert and Mudry}(2011)}]{Neupert:2011aa}
\bibinfo{author}{\bibfnamefont{C.~C.} \bibnamefont{Titus~Neupert},
  \bibfnamefont{Luiz~Santos}} \bibnamefont{and}
  \bibinfo{author}{\bibfnamefont{C.}~\bibnamefont{Mudry}},
  \bibinfo{journal}{Physical Review Letters} \textbf{\bibinfo{volume}{106}}
  (\bibinfo{year}{2011}).

\bibitem[{\citenamefont{Trescher and Bergholtz}(2012)}]{Trescher:2012aa}
\bibinfo{author}{\bibfnamefont{M.}~\bibnamefont{Trescher}} \bibnamefont{and}
  \bibinfo{author}{\bibfnamefont{E.~J.} \bibnamefont{Bergholtz}},
  \bibinfo{journal}{Physical Review B} \textbf{\bibinfo{volume}{86}}
  (\bibinfo{year}{2012}).

\bibitem[{\citenamefont{Ishikawa and Matsuyama}(1987)}]{ISHIKAWA1987523}
\bibinfo{author}{\bibfnamefont{K.}~\bibnamefont{Ishikawa}} \bibnamefont{and}
  \bibinfo{author}{\bibfnamefont{T.}~\bibnamefont{Matsuyama}},
  \bibinfo{journal}{Nuclear Physics B} \textbf{\bibinfo{volume}{280}},
  \bibinfo{pages}{523 } (\bibinfo{year}{1987}), ISSN \bibinfo{issn}{0550-3213},
  \urlprefix\url{http://www.sciencedirect.com/science/article/pii/055032138790160X}.

\bibitem[{\citenamefont{Golterman et~al.}(1993)\citenamefont{Golterman, Jansen,
  and Kaplan}}]{Golterman1992ub}
\bibinfo{author}{\bibfnamefont{M.~F.~L.} \bibnamefont{Golterman}},
  \bibinfo{author}{\bibfnamefont{K.}~\bibnamefont{Jansen}}, \bibnamefont{and}
  \bibinfo{author}{\bibfnamefont{D.~B.} \bibnamefont{Kaplan}},
  \bibinfo{journal}{Phys. Lett.} \textbf{\bibinfo{volume}{B301}},
  \bibinfo{pages}{219} (\bibinfo{year}{1993}), \eprint{hep-lat/9209003}.

\bibitem[{\citenamefont{Volovik}(2009)}]{Volovik_2009}
\bibinfo{author}{\bibfnamefont{G.~E.} \bibnamefont{Volovik}},
  \emph{\bibinfo{title}{The Universe in a Helium Droplet}}
  (\bibinfo{publisher}{Oxford University Press}, \bibinfo{year}{2009}),
  \urlprefix\url{https://doi.org/10.1093%2Facprof%3Aoso%2F9780199564842.001.0001}.

\bibitem[{\citenamefont{Zubkov}(2016)}]{ZUBKOV2016298}
\bibinfo{author}{\bibfnamefont{M.}~\bibnamefont{Zubkov}},
  \bibinfo{journal}{Annals of Physics} \textbf{\bibinfo{volume}{373}},
  \bibinfo{pages}{298} (\bibinfo{year}{2016}), ISSN \bibinfo{issn}{0003-4916},
  \urlprefix\url{https://www.sciencedirect.com/science/article/pii/S0003491616301130}.

\bibitem[{\citenamefont{Nakahara}(2003)}]{nakahara2003geometry}
\bibinfo{author}{\bibfnamefont{M.}~\bibnamefont{Nakahara}},
  \emph{\bibinfo{title}{Geometry, topology and physics}}
  (\bibinfo{publisher}{CRC Press}, \bibinfo{year}{2003}).

\bibitem[{\citenamefont{Zhang et~al.}(2013)\citenamefont{Zhang, MacDonald, and
  Mele}}]{ZhangMacDonald2013}
\bibinfo{author}{\bibfnamefont{F.}~\bibnamefont{Zhang}},
  \bibinfo{author}{\bibfnamefont{A.~H.} \bibnamefont{MacDonald}},
  \bibnamefont{and} \bibinfo{author}{\bibfnamefont{E.~J.} \bibnamefont{Mele}},
  \bibinfo{journal}{Proceedings of the National Academy of Sciences}
  \textbf{\bibinfo{volume}{110}}, \bibinfo{pages}{10546}
  (\bibinfo{year}{2013}),
  \eprint{https://www.pnas.org/doi/pdf/10.1073/pnas.1308853110},
  \urlprefix\url{https://www.pnas.org/doi/abs/10.1073/pnas.1308853110}.

\bibitem[{\citenamefont{Hashimoto and Kimura}(2016)}]{Hashimoto2016}
\bibinfo{author}{\bibfnamefont{K.}~\bibnamefont{Hashimoto}} \bibnamefont{and}
  \bibinfo{author}{\bibfnamefont{T.}~\bibnamefont{Kimura}},
  \bibinfo{journal}{Progress of Theoretical and Experimental Physics}
  \textbf{\bibinfo{volume}{2016}} (\bibinfo{year}{2016}), ISSN
  \bibinfo{issn}{2050-3911}, \bibinfo{note}{013B04},
  \eprint{https://academic.oup.com/ptep/article-pdf/2016/1/013B04/9620604/ptv181.pdf},
  \urlprefix\url{https://doi.org/10.1093/ptep/ptv181}.

\bibitem[{\citenamefont{Wilson}(1977)}]{Wilson_1977}
\bibinfo{author}{\bibfnamefont{K.~G.} \bibnamefont{Wilson}}, in
  \emph{\bibinfo{booktitle}{New Phenomena in Subnuclear Physics}}
  (\bibinfo{publisher}{Springer {US}}, \bibinfo{year}{1977}), pp.
  \bibinfo{pages}{69--142}.

\bibitem[{\citenamefont{Qi et~al.}(2006)\citenamefont{Qi, Wu, and
  Zhang}}]{QWZ2006}
\bibinfo{author}{\bibfnamefont{X.-L.} \bibnamefont{Qi}},
  \bibinfo{author}{\bibfnamefont{Y.-S.} \bibnamefont{Wu}}, \bibnamefont{and}
  \bibinfo{author}{\bibfnamefont{S.-C.} \bibnamefont{Zhang}},
  \bibinfo{journal}{Phys. Rev. B} \textbf{\bibinfo{volume}{74}},
  \bibinfo{pages}{085308} (\bibinfo{year}{2006}),
  \urlprefix\url{https://link.aps.org/doi/10.1103/PhysRevB.74.085308}.

\bibitem[{\citenamefont{Koshino and McCann}(2009)}]{KoshinoMacCann2009}
\bibinfo{author}{\bibfnamefont{M.}~\bibnamefont{Koshino}} \bibnamefont{and}
  \bibinfo{author}{\bibfnamefont{E.}~\bibnamefont{McCann}},
  \bibinfo{journal}{Phys. Rev. B} \textbf{\bibinfo{volume}{80}},
  \bibinfo{pages}{165409} (\bibinfo{year}{2009}),
  \urlprefix\url{https://link.aps.org/doi/10.1103/PhysRevB.80.165409}.

\end{thebibliography}

\end{document}